\documentclass[twocolumn,10pt]{article}

\usepackage[margin=1in]{geometry}
\usepackage{times}
\usepackage{microtype}
\usepackage{hyperref}
\usepackage{booktabs}
\usepackage{array}
\usepackage{tabularx}
\usepackage{enumitem}
\usepackage{titlesec}
\usepackage{authblk}
\usepackage{abstract}
\usepackage{parskip}
\usepackage{footnote}
\usepackage{url}

\hypersetup{
  colorlinks=true,
  linkcolor=black,
  urlcolor=blue,
  citecolor=black
}

\titleformat{\section}{\normalfont\bfseries}{\thesection}{1em}{}
\titleformat{\subsection}{\normalfont\bfseries\itshape}{\thesubsection}{1em}{}

\renewenvironment{abstract}
  {\small\quotation\noindent\textbf{Abstract.}\enspace}
  {\endquotation}

\title{\textbf{The End of Human Judgment in the Kill Chain? Relocating Initiative and Interpretation with Agentic AI}}

\author[1,2,3]{Jovana Davidovic}
\affil[1]{Senior Researcher, Peace Research Institute Oslo}
\affil[2]{Associate Professor, University of Iowa}
\affil[3]{Chief Ethics Officer, BABL AI}

\date{}

\begin{document}

\maketitle

\begin{abstract}
Large language model-based agents are increasingly being integrated into core battlefield functions, including intelligence analysis, data fusion, and battlefield management. This paper argues that the very features that make such agents operationally attractive, namely their capacity for initiative, interpretation, their goal-directedness, and dynamic memory, are the same features that render context-appropriate human judgment and control substantively ineffectual in those parts of the kill chain where agents operate. Drawing on specific use cases, the paper argues that by relocating initiative and interpretation, LLM-based agents displace human decision-making in ways that makes their use incompatible with the requirement of human judgment and control which is central to existing governance frameworks, like those proposed by the GGE-CCW and REAIM. The paper concludes that a subset of agentic AI applications, particularly those deployed for data fusion and battle management in lethal contexts, cannot be used justifiably on the battlefield under current and foreseeable conditions, and proposes two ways for the international governance community to respond to this challenge.
\end{abstract}

\section*{Introduction}

AI agents are increasingly promoted as force multipliers in military operations, especially for intelligence analysis, data fusion, and battle management. This paper examines specific such use cases and argues that, in lethal contexts, some forms of agentic AI, specifically, LLM-based agents, are not compatible with the requirement of context-appropriate human judgment and control and present significant risks that make it unlikely that such tools can be used justifiably on the battlefield for the foreseeable future [22, 7].

Drawing on specific examples of proposed and emerging military applications, the paper argues that the very features that make LLM-based agentic AI operationally attractive (features like initiative, interpretation, goal oriented-ness, and dynamic memory) also undermine meaningful human engagement. In these types of agents, human involvement is increasingly displaced, creating distinctive ethical risks that go beyond those posed by earlier AI-enabled tools.

The main claim in this paper is that while some agentic AI applications may remain compatible with human control, there is a \textit{subset} of agentic systems which are fundamentally unsuitable for use in the kill chain. By reallocating initiative and interpretive authority to the system itself, these agents reconfigure the human role in ways that make their ethical deployment in lethal settings unattainable, if context-appropriate human judgment and control is necessary for ethical deployment.

The paper is organized in six sections. In the next section (section II), I discuss a specific use case, which focuses on data fusion on the battlefield. While it is unclear how far use cases such as this one have gotten in practice, I provide a number of reasons to believe that LLM-based agents are going to be increasingly used by militaries for these types of capabilities. In this section, I also say a little bit about what LLM-based agents are and what types of agents I have in mind in this paper. In section III, I identify four features that make such agents attractive for exact capabilities discussed in section II, namely data fusion, intelligence analysis, and battlefield management. These four features include initiative, interpretation, goal-oriented behavior, and dynamic memory. I discuss and describe these features and show why they make LLM-based agents attractive for the above capabilities. In section IV, I argue that these four features are the exact features that make the LLM-based agents particularly risky and I argue that these four features which enable agents to outperform ordinary AI (with respect to some capabilities and in some parts of the kill chain) \textit{render human judgment in those parts of the kill chain substantively ineffectual} (or not-effectual enough to do the work we need human judgment to do). Additionally, in this section, I stress that all the key policy conversations about governance of AI-enabled weapons systems and autonomous weapons systems (namely the CCW and REAIM processes) have been calling for human judgment in the deployment of a weapon system (or pre-deployment). I argue that the use of agents of the above type (for data fusion) render human judgment in that part of the kill chain ineffectual and thus incompatible with those policy proposals. In section V, I examine potential objections to my argument, and finally in section VI, I consider two ways forward for the international governance and policy community. Specifically, I consider a potential ban on the use of LLM-based agents for data fusion and battlefield management, and after rejecting it as unrealistic, I turn to what I consider to be the next best strategy for mitigating the significant risks that these agents pose in these capabilities on the battlefield.

\section*{II. LLM-agents on the battlefield}

Speed is one of the primary drivers of success on the battlefield. A central bottleneck that puts an upper limit on speed on the battlefield is data processing, analysis, and fusion. Data on the battlefield comes in from a variety of sources---sensor data (e.g.\ video or IR feed from a drone), geospatial data, human intelligence, signal intelligence (e.g.\ intercepted communications), etc. With an increase in the number of ISR drones, satellite data, and ability to intercept and process online communications, the sheer volume of data presents a serious problem for gathering useful and actionable data on the battlefield. AI has been widely used for a long time now for solving just such a problem. Consider for example, Anduril's Lattice for Command and Control (C2) [1]. Anduril describes it as an ``AI-powered battle-management platform built to accelerate complex kill chains,'' by integrating ``thousands of sensors and effectors to turn data into decisions at scales and speeds beyond human capacity.'' [1] In short, Lattice acts like a central brain or operating system which connects different sensors and devices (e.g.\ drones, cameras, radar, surveillance towers) into one unified picture. By fusing that data together, it allows the operator a unified single pane view of what's happening. Centrally, for our purposes, Lattice C2 engages in data translation and normalization and ``it breaks down traditional data silos by integrating and normalizing heterogenous data in real-time, serving as a translator that can speak the unique language of each service, platform, robot, etc.'' [1] Lattice uses AI to identify and filter out the noise and to highlight what is important, by ``intelligently and automatically classifying, evaluating, and prioritizing data to get the right data to the right persons at the right time in limited bandwidth scenarios.'' [1] Lattice C2 can also recommend actions, for example to deploy a drone to collect further information in cases where incoming data is degraded or confidence levels are low. Finally, as a C2 system Lattice can turn decisions into actions tasking any assets it is connected to including manned, unmanned, or even autonomous tools and weapon systems [1].

While Lattice uses AI and deep learning models, it is not clear, and there is no public evidence to think, that it uses an \textit{LLM-based} agent or agents at its heart. In general, it is unclear how far use cases such as this one have gotten, however there are really good reasons to think that LLM-based agents are going to be increasingly used by militaries for the above types of capabilities. In what follows I will say much more about what those reasons are, but to be able to provide these reasons I quickly want to describe what I have in mind when I say `LLM-based agents,' as this is the type of agents that I think are most likely to be the ones that will be put in place to perform functions that tools like Lattice C2 perform.

An LLM-based agent combines a large language model (LLM) with additional components so it can ``reason,'' plan, and act [24]. In such an agent the LLM serves as a brain, understanding goals and generating strategies, and it uses external tools (such as API calls, databases, search, code execution) to interact with the world. More importantly, an orchestration layer powered by such an LLM manages the functional loop that agents use to perform its function. Namely, the LLM's role is to perceive context and ``understand'' the intent of the operators, plan which steps to take to accomplish the goal in that context, and then use the above tools to reason and act, observing the results, assessing those results, and repeating the sequence if the goal has not been achieved. Agents of this type use LLM-based reasoning frameworks (like ReAct or chain-of-thought) to decide when to think and when to act. They also maintain a memory of past steps, and they can break complicated, multi-step tasks into manageable subtasks (partly what I mean when I say they can ``understand'' intention of operators). This architecture enables autonomous problem solving well beyond standalone LLMs and beyond ordinary (non-LLM) agents.

While there are many ways an LLM-based agent could be organized, it is worth quickly considering what an LLM-based orchestration layer would look like in practice for an agent performing data fusion. The architecture would most likely consist of multiple specialized nodes or sub-calls, each handling a specific task like normalization, filtering, or contextual inference. Importantly, the situational context itself could be dynamically generated by the LLM (e.g.\ ``we are now tracking trucks in an off-grid area'') and passed as a natural-language description to downstream nodes. A normalization node, for instance, would receive that context and use it to intelligently reconcile conflicting speed readings across sensor types, recognizing, say, that a radar return showing 45 mph and a visual tracker reporting 38 mph for the same vehicle are likely measuring the same object under different latency and angle conditions, rather than flagging them as separate contacts, something that would otherwise require hardcoded sensor-fusion heuristics tuned to each specific pairing.

There are a number of reasons to think that these types of agents are or will be utilized in the battlefield. First, these are the types of agents that are most likely to be well suited to solve capabilities bottlenecks for data fusion and intelligence analysis. I will address this point (about how well suited they are and why) in much more detail in the following section. Second, industry, including the defense industry, is already using these types of agents for workflows in development, design, testing, governance, and monitoring of AI models [5]. In other words, LLM-based agents are already playing a role in the lifecycles of warfighting AI models. Third, non-defense industry widely uses LLM agents for similar (data fusion) tasks [5, 17].\footnote{For example, consider banking and fraud detection: Fusing a customer's real-time transaction stream with historical spending behavior, device fingerprint data, geolocation pings, and third-party credit bureau signals to produce a unified risk picture of a single account holder at a given moment.} Fourth, the Departments and Ministries of Defense around the world have explicitly said they are planning to use agents in this way and are investing in such solutions. For example the January 9th, 2026 AI Strategy for the U.S.\ Department of Defense explicitly lists developing and using agentic networks for warfighting, saying that the DoD intends on ``unleashing AI agent development and experimentation for AI-enabled battle management and decision support, from campaign planning to kill chain execution.'' [23] A few days later the Defense Innovation Unit announced a 100 million dollar prize for orchestrators for tasking autonomous vehicles, saying ``[w]e want orchestrator technologies that allow humans to work the way they already command---through plain language that expresses desired effects, constraints, timing, and priorities---not by clicking through menus or programming behaviors.'' [10, 21] An obvious (and maybe only) solution for using plain language for tasking numerous autonomous tools are LLM-based agents. In addition, the recent (March 2026) dust up between Anthropic and the DoD as well as the use of LLMs in the 2026 war on Iran has unearthed the wide extent of use of LLMs and intended plans of use of LLMs by the U.S.\ military [3, 15, 18, 19]. All of this is simply to say---LLM agents for data fusion and battlefield management are being built, tested, and/or sold \textit{now}. Additionally, and likely more importantly, there are good reasons to think that LLM-based agents are the best solution for data fusion and battlefield management. I turn to this argument next.

\textit{Agentic LLM-based} systems \textit{in contrast} can actively manage sensors, they can task a different sensor if data from one is degraded, they can prioritize sensor data streams, they can track uncertainty and confidence over time, they can remember potentially where a sensor worked better and which sensor or types of data to trust or confidence levels to accept in which environments. Agents of this type can resolve conflict by querying additional sources as well as change how they fuse data from multiple sources based on mission context. After all that is exactly how LLM-based agents work, by translating particular instructions or intent into a planned set of tasks and reiterating on data collection and prioritization until the goal has been reached.

Simply, LLM-based agents seem better suited to the capabilities like data fusion, and battlefield management. This is partly because LLMs can interpret goals expressed in natural language on the fly, without those goals being pre-programmed. Consider for example a hostage-in-a-truck example: a commander can say in plain language ``we think there's a hostage on one of these trucks, figure out which one,'' and an LLM agent can reason across all available sensor data (e.g., noticing that three cameras picked up a small amount of blue consistent with the hostage's clothing) to generate a probabilistic answer with reasoning. This kind of flexible, contextual inference over heterogeneous, potentially conflicting inputs was/is essentially out of reach for hardcoded or even ML-trained pre-LLM agentic systems.

With this general description of why LLM-based agents are likely much better suited to serve the command and control and data fusion capability, I can now turn to those (four) \textit{main features} of LLM-based agents that provide/explain \textit{why} such systems have the above benefits over legacy systems (including legacy AI systems). As mentioned, these features include initiative, interpretation, goal-orientedness, and dynamic memory.

\section*{III. Key battlefield-capability-enabling features of LLM-based agents}

In this section I turn to four key features that make LLM agents such as the ones described above attractive for exact capabilities discussed above, namely data fusion, intelligence analysis, and battlefield management. These include agent's \textit{initiative}, reliance on and ability to do \textit{interpretation}, \textit{goal-oriented} behavior, and \textit{dynamic memory}. Below I discuss and describe each of these four features and show why they make LLM-based agents attractive for the above functions/capabilities (data fusion, intelligence analysis, and battlefield management---with a focus on data fusion). Before turning to each of these features individually, it is worth elaborating in more general terms why LLM-based agents are advantageous over legacy or ordinary/conventional AI systems for these battlefield capabilities.

Consider again the case alluded to before where a battle station receives feeds from hundreds or more sources and dozens of sensor types. Each of the sensor types has a different update rate or reliability---maybe performing better in some environments, but not others. Information coming in from all these sources would often obviously not only have different confidence levels, but also it might be conflicting or incomplete. This is a common problem on the battlefield, and one that Lattice C2 is developed for, namely to sift through enormous amounts of data and decide what to trust, what data is best, which data to discard, recognize when some data is degraded, etc.

Imagine now having an ``ordinary'' deep learning AI at its heart. Such a system would most likely have to assume static, well-curated inputs from each sensor, it would struggle with uncertainties and temporal reasoning, as well as either struggle with or not at all be capable of deciding what data to seek next. A command-and-control system built on a traditional agent architecture has a centralized ``brain'' which receives data, makes decisions, and sends outputs either to update internal state (databases, memory) or to external actuators (drones, satellites, displays). Inputs and outputs for a traditional agent, however, must be spatially standardized (as in, everything needs to share a common coordinate system, like GPS/GNS coordinates) and likely must have consistent labeling conventions so the system could treat data from different sensors as referring to the same object or location. This is the baseline for traditional agents and ordinary AI acting as the brain for a system like the Lattice C2: a structured, deterministic pipeline rather than anything reasoning flexibly. A central limitation of pre-LLM agentic systems is that the goals used to filter and prioritize incoming data had to be explicitly programmed (or ML programmed) in advance. The system could be set to alert when something crosses a perimeter, when a battery level drops, or when an object appears, but those conditions had to be anticipated and coded before deployment. True dynamic goal setting was not effectively possible (e.g., a commander saying mid-operation ``I now care about finding a hostage''). Data fusion, similarly, is likely rudimentary in such a system: rather than weighing conflicting evidence across sensors intelligently, the system probably just displays the highest-confidence sensor output, perhaps with a count of how many other sensors agreed. It is worth noting that reinforcement learning-based agents or recommendation-system-style models could, in theory, have been trained to do sensor fusion and prioritization. You could imagine training a model on simulated commander behavior, what outputs commanders flagged as important, and building a system that learned to surface the most relevant signals. However, this is unlikely to have been the case for most systems trying to act in this capacity, because it would have required enormous amounts of clean, standardized training data from diverse sources, data that likely doesn't exist in sufficient quantity.

\begin{enumerate}[label=\alph*.]
\item \textbf{Initiative (including self-halting) and dynamic tasking:} LLM-based agents can do things that would otherwise require human input and they can decide when to seek human input. They can dynamically task sensors and systems. Once given a particular mission parameter or task they can make a plan how to collect the data they need to achieve that task, but also if they don't feel confident that sufficient data has been gathered (for example to recommend object on the ground as a potential target, or to recommend this rather than that asset for engaging such a target) the agent can task other sensors, switch to another type of analysis, refer back to previous experiences with similar targets or in this environment, etc. Simply put, the system has a lot of initiative, and it doesn't need to go ``back'' to the human operator to task tools that it needs to get to confidences it needs to feel that the task has been completed. In addition, since the system can process massive amounts of data to make tasking choices, it likely is better than the humans in making those types of tasking choices, making it also much faster (and speed wins wars).

\item \textbf{Interpretation:} LLM-based agents can interpret and ``reason'' over unstructured data (like raw sensor feeds, radio transcripts, intelligence reports, imagery metadata, operator notes) and reason across all of it simultaneously, something traditional systems struggle to do. Rather than requiring data to be pre-formatted into rigid schemas and labels, an LLM-based agent can read a fragmented drone observation, cross-reference it against a partially redacted SIGINT report, and a commander's voice-transcribed intent, then surface a coherent operational picture. Take the example we started from (Anduril's Lattice): today (to the best of our knowledge) it fuses information from various sources and sensor tracks into a common operating picture using structured pipelines. If Lattice were LLM-powered, it could go further, an agent might read an unstructured after-action report noting ``enemy vehicles were observed staging near the northern tree line around dusk,'' correlate that against thermal sensor anomalies from the night prior, and autonomously flag a probable pattern-of-life threat without a human analyst manually tagging each data point first. It is this type of interpreting ability that makes LLM-agents particularly suitable for these types of problems. In addition, and maybe even more importantly, such agents can reason across different types of data and in the context of significant ambiguity. An LLM-based battlefield agent could reconcile a UAV feed that shows \textit{4} vehicles with a ground patrol report describing ``a \textit{small} convoy'' and a logistics intercept referencing fuel requests for \textit{6} trucks, and reason that the discrepancy itself is tactically significant, rather than simply averaging or discarding conflicting inputs. That kind of inferential data fusion over noisy, incomplete, and heterogeneous battlefield data, is where LLM agents meaningfully improve outcomes beyond legacy (including ``old'' AI) fusion architectures.

\item \textbf{Goal Oriented Behavior:} LLM-based agents can create sub-goals and prioritize them dynamically and in varying contexts. LLM-based agents don't just pursue a single top-level objective, or intention from a commander directly. Instead, they decompose a goal into sub-goals and they can reprioritize each element (or step in the plan) if and when the context shifts. In a battlefield management scenario, a Lattice-like LLM agent which is given the directive ``maintain air superiority over sector 7'' might autonomously generate sub-goals around ISR coverage, threat identification, and intercept coordination and then dynamically reorder those priorities the moment a new radar track appears, without waiting for a human to rewrite the task queue. This is fundamentally different from classical planning systems, which require pre-enumerated goal hierarchies. An LLM-based agent can recognize that a context change (say, a comms blackout or an unexpected ground incursion) invalidates certain sub-goals entirely requiring either a restructuring or brand new sub-goals, by ``reasoning'' about what matters \textit{now} versus what the original plan assumed. In that way this relates to its interpretive ability discussed above, and to dynamic memory discussed below. Translating goals into sub-goals and ``understanding'' how both the breakup of goals into sub-goals and their prioritization are related to the context within which the goal was set allows for the changes in prioritizations and sub-goals relative to not just the intent, but context.

\item \textbf{Dynamic Memory:} LLM-based agents, discussed here, can update and reference facts about the world dynamically across time and tasks integrating it with past goals and observations. This too is related to the above features, but it is centrally important because LLM-based agents can maintain and update their dynamic memory---they can log past observations and past decisions, and keep track of evolving world states, and most importantly they can then actively draw on that history when reasoning about new inputs. A Lattice-like LLM agent might remember that a particular grid sector was flagged as a low-threat six hours ago, but upon receiving new radar data, recall that a similar pattern preceded an ambush in a prior engagement and elevate its threat assessment accordingly. This goes beyond simple data logging; an agent such as this would be using past context and observations and decisions for making new inferences, in other words, it could connect a current observation to a chain of prior goals and outcomes.
\end{enumerate}

\begin{table}[h]
\small
\begin{tabularx}{\columnwidth}{>{\bfseries\itshape}l X}
\toprule
Feature & Key Benefits \\
\midrule
Initiative & Ability to do things that would otherwise require human input and ability to decide when to seek further input; dynamic tasking of sensors and devices \\[4pt]
Interpretation & Ability to interpret and ``reason'' over unstructured data; ability to ``respond'' to different contexts and interpret orchestration instructions \\[4pt]
Goal-orientedness & Goals translated into sub-goals and tasks/steps; sub-goals can be created and prioritized dynamically and in varying contexts \\[4pt]
Dynamic memory & Ability to update and reference facts about the world dynamically across time and tasks integrating it with past goals and observations \\
\bottomrule
\end{tabularx}
\caption{Key battlefield-capability-enabling features of LLM-based agents}
\end{table}

These features are related and are by no means the only important or risk-grounding features of LLM-based agents [16]. Instead, any agent that does have these features, I will argue, presents a significant, potentially unmitigable, risk in a kill chain (given certain assumptions about what types of governance and oversight we want over the kill chain). One example of how they (the features) intersect or interact in producing the key benefits discussed above includes the fact that LLMs can interpret and re-interpret particular instructions and information in different ways given the change in context---and that change in context can be driven by the systems' dynamic memory. Obviously, the key benefits emerge from these features interacting, but they are, as I mentioned, mostly individuated here so as to explain how they contribute to unique risks, and specifically how they contribute to relocating normative and epistemic authority away from humans to machines, thus undermining meaningful human oversight and control. In the next section I turn to my main argument, namely that the main features of LLM-based agents discussed above ground significant ethical risks and that they make context-appropriate human judgment over the kill chain impossible. It is worth noting here that while I think there are many ethical and legal risks that would emerge if and when LLM agents are used for data fusion and battlefield management, I focus here on how such agents undermine human oversight and meaningful engagement with the system.

\section*{IV. Key features of LLM-based agents and the significant (new) risks on the battlefield}

In what follows I argue for the main claim of this paper, namely that \textit{the exact features that enable LLM agents to outperform ordinary AI (with respect to some capabilities and in some parts of the kill chain) also render human judgment in those parts of the kill chain substantively ineffectual (or not-effectual enough to do the work the international community has been asking that human judgment perform).} They do so by relocating initiative and interpretation from humans to machines. Each of the above key features grounds a new potential and significant ethical risk in the battlefield [6]. While, in what follows, I examine a range of those risks, the primary focus here is on the reallocation of normative and epistemic authority and what that means for asserting that humans can provide meaningful oversight, involvement, engagement, judgment, or control over the system [14].

\begin{enumerate}[label=\alph*.]
\item Let's begin with initiative. As mentioned above, LLM-based agents can do things that would otherwise require human input and they can decide when to seek human input. This in turn clearly redistributes (reallocates) normative and epistemic authority. By normative authority I mean the right and power to make morally and legally salient choices (i.e.\ the right and power to make decisions that create obligations, assign responsibilities, determine what ought to be done in a particular situation, and result in morally and legally salient outcomes). By epistemic authority I mean the right and obligation to form consequential and appropriate beliefs and make judgments.\footnote{It is worth noting here that this use of `epistemic authority' is not a standard philosophical one (which focuses on someone whose beliefs you have good reason to defer to because they're more reliable or expert). My use here focuses on the \textit{obligation} to form beliefs, not just the right.} Jointly this phrase ``normative and epistemic authority,'' is meant to capture \textit{the right and power (with concomitant obligations) to make morally and legally salient decisions over epistemically appropriately justified beliefs.}

As discussed above and regarding initiative, in ordinary AI systems in these sorts of roles the system is pre-programmed, whether by explicit rules or as a ML system, in deterministic ways to seek input from a human at certain thresholds, or confidence levels, at certain places in the decision-chain, etc. On the other hand, systems that use LLMs in the orchestration layer depend on those LLMs to decide when to seek human input. The agent can task further sensors, or drones and it can dynamically change when to seek input based on its memory. For example, recognizing that last time when a system encountered this type of an enemy vehicle in this context the tree shade altered its color, the system can attempt to proceed with lower confidence levels or deploy another type of sensor that can confirm correct target is identified. In other words, the system can take initiative to make sure it accomplishes its task without going back to the human. In fact, even in cases when the orchestration level might have explicit lines of code for when to seek human input, interpreting what those lines of code require would fall to an LLM. All of this is simply to illustrate what I mean when I say that the feature of initiative redistributes normative and epistemic authority. The fact that the agent can make decisions regarding when to collect further evidence, or when to call on a human, how to assess confidence levels against past experiences and similar all points to the fact that \textit{key decisions that have significant ethical consequences and epistemic consequences} are left up to the agent.

\item With respect to interpretation, there is a range of issues, in addition to the above mentioned fact that leaving interpretation of rules (in the orchestration layer) to an LLM redistributes epistemic authority almost by definition. The key other issue with interpretation being left in large part to an LLM is that it introduces a new level of opacity that can affect user's ability to make justified judgments and therefore our ability to attribute responsibility for outcomes. While opacity is most certainly a problem with any deep learning neural network and has also been a problem for legacy AI tools used for data fusion, intelligence analysis, and battlefield management, the introduction of stochastic LLMs into these capabilities and the assignment of interpretation of rules to LLMs makes opacity an even more significant problem. All in all, the reliance on LLMs in the orchestration layer of agents redistributes epistemic authority and introduces a new type of opacity. These two facts jointly make responsibility attribution for bad outcomes significantly more difficult if not impossible. I will say more about this shortly, but ability to attribute responsibility for a particular outcome is one of the key requirements of most policies intended to govern AI-enabled kill chains and weapon systems.

\item With respect to goal-orientedness, there are significant worries about the system's ability to preserve the intent of the commander/operator. As mentioned above, for each command or instruction, LLM-based agents create sub-goals and prioritize those sub-goals dynamically and in accord with varying contexts. LLM-based agents don't just pursue a single top-level objective, or intention from a commander/operator directly. Instead, they decompose a goal into sub-goals, and they reprioritize each element (or step in the plan) if and when the context shifts. This in turn means that the LLM is responsible for \textit{re-conceptualizing} the intent---both the commanders'/operators' intent and even the developer's intent. As I will argue shortly, preservation of intent is once again one of the key requirements of most policies relying on human judgment for governing warfighting AI.

\item Finally, with respect to the feature of dynamic memory, the reliance on it can reshape the context against which intent in interpreted and against which judgments are made---thus significantly affecting outcomes. As a reminder, LLM-based agents can update and reference facts about the world dynamically across time and tasks, integrating it with past goals and observations (i.e.\ they have a dynamic memory). That in turn means that the agent's dynamic memory can significantly affect how the system prioritizes tasks or even how it organizes various sub-goals, it can affect what tools it tasks or which confidence levels it deems acceptable. It follows then that such agents can respond differently to circumstances that might appear normatively identical to the observer based on a wealth of contextual and past observation information. This complicates assessment of outcomes---i.e.\ did the agent simply know more than the observer or operator, or did the system act by mistake? Dynamically reshaping the context against which intentions or commands are interpreted clearly leads to displacement of both normative and epistemic authority and potentially complicates our ability to identify mistakes.
\end{enumerate}

\begin{table}[h]
\small
\begin{tabularx}{\columnwidth}{>{\bfseries\itshape}l X}
\toprule
Feature & Key Risks \\
\midrule
Initiative & Displacement of normative and epistemic authority \\[4pt]
Interpretation & Opacity and displacement of epistemic authority \\[4pt]
Goal-orientedness & Re-conceptualization of human intent (thus displacing normative authority) \\[4pt]
Dynamic memory & Displacement of epistemic and normative authority by dynamic alteration to the background context against which beliefs are formed and morally salient decisions are made \\
\bottomrule
\end{tabularx}
\caption{Key risks associated with LLM-based agent features}
\end{table}

So why is all this a problem? After all, AI tools of all kinds and complex systems of all kinds might share some of these risks---like opacity, complicating responsibility attribution, and similar. The reason the above matter and present a risk that likely \textit{cannot be mitigated} has to do with the fact that so-called ``context appropriate human judgement and control'' cannot be met for most LLM-based agents in these capabilities (e.g.\ data fusion in the kill chain or for targeting). I have argued above that the same features that make LLM-agents attractive in this part of the kill chain, also render human judgment in those parts of the kill chain \textit{substantively ineffectual} (or not-effectual enough to do the work the international community has been asking that human judgment perform) [14].

This than raises a key question---why do ``context appropriate human judgment and control'' matter and what does the condition for such human judgment and control require of us and of AI systems? In other words, we need to ask: what is the role of ``human judgement'' or ``human engagement'' in the proposed policies and why do policy makers and scholars insist on human judgment in the kill chain? Answering that question informs \textit{what shape} such human judgment needs to take and \textit{where} in the kill chain such human judgement needs to be present.

This topic has been discussed in great lengths elsewhere, thus here I only briefly review the main points of those discussions. First, it is worth noting that all current policy proposals for governing AI-enabled weapons call for some type of human judgment in the lifecycle of the weapon \textit{and} in the so-called kill chain. Second, there is a large spectrum of views regarding where and how human judgment ought to be exerted in the kill chain. Part of the reason for this variability on operationalizing human judgment comes from variability with respect to the \textit{purposes} for which policy makers and scholars insist on human judgment in the lifecycle and kill chain. In other words, different purposes behind calls for human judgment result unsurprisingly in different views on where and when human judgment matters the most in the kill chain and how it should be exercised. These purposes (behind the calls for human judgment and/or control) include safety, responsibility attribution, human dignity, and institutional stability [8, 9]. In other words, some scholars and policy makers think that human judgment in the kill chain will make the weapons safer, others think that a human should always remain in control of a system so that we can always trace back responsibility for bad outcomes to a human, and yet others think only humans can really implement key policies and rules (like international humanitarian law rules) because those rules were written with human executors in mind. In addition to a wide range of views regarding why we want humans in the kill chain, there is also a spectrum of views on the \textit{type} of human engagement/involvement called for (e.g.\ control, judgment, deliberation, preserving intent, or preserving values) [8, 9]. For example, some insist on human control over the system (for example so we can assign responsibility) and others just want preservation of human intent (to assure responsibility and increase control and potentially safety of systems). But, while there is a range of views on why we want humans in the kill chain and what type of oversight or control such humans should exerts, there is wide agreement that humans should stay in the kill chain. In other words, the fact \textit{that} some form of context-appropriate human judgment \textit{is required} in the lifecycle of AI-weapons \textit{and} in the kill chain is agreed upon among most policy makers and scholars even when such agreement is weak (as in, even when it is not clear what the agreement is over in practice).

To illustrate, consider the main ongoing multilateral policy discussions; namely, the conversations around the rolling text by the Group of Governmental Experts on the Convention on Certain Conventional Weapons (GGE-CCW) and the Strategic Priorities Document that emerged from the REAIM process [13, 20, 22].

The GC-REAIM Strategic Priorities document starts by thinking of the entire lifecycle of an AI system as the appropriate landscape for governance [13]. The report then argues that a key concern for governance of AI weapons and kill chains ``lies in the potential erosion of human judgment in decisions around the use of force.'' [13] Most centrally, the document asserts that ``human moral agency, the capacity to make contextual, evaluative judgments about `right and wrong' while retaining responsibility for the actions and choices pursued'' remains central to the ethical and lawful use of AI systems in the military domain. [13]

Similarly, the GGE-CCW's work, which has probably been the most significant multilateral process on the issue of use of autonomous systems for warfighting (or LAWS---lethal autonomous weapon systems), shares this focus on the importance of human agency and human judgment. The current rolling text has been the object of negotiations for years and has undergone many iterations, all of which included reference to the importance of human judgment or control. The most recent version stresses that the ``context-appropriate human judgement and control is needed to ensure the use and effects of LAWS are in compliance with international law, in particular IHL, including the principles and requirements of distinction, proportionality and precautions in attack.'' [7]

While there has been a lot of back and forth around the exact language, there is wide agreement across international, multilateral, and domestic conversations that something akin to ``context-appropriate human judgment'' is required for justified warfighting. There has also been growing agreement that such human judgment can take various forms, including exerting such judgment prior to deployment (leaving even space for lethal autonomous weapon systems).

Even with the significant variability in the calls for human judgment in the kill chain, one thing is for certain, everyone agrees that human judgment and human agency in the kill chain matter [2]. What such judgement or agency looks like is, as we have seen, a matter of considerable disagreement. But there is good reason to think that there is a bare minimum or a lower threshold of human agency in the kill chain that must be met (regardless of whether we want human judgment, or control over the system). So what could this bare minimum look like? I take it that whatever the bare minimum is, it is such that human agency or human judgment can be referenced back \textit{as a key part of an explanation} and \textit{as a key part of the justification} for a particular outcome. Of course, what constitutes an adequate explanation, and an adequate justification of an outcome are themselves deeply contested questions in both philosophy and law. What I have in mind here is something quite modest: an ordinary-language account of what we would point to as the key drivers of a particular outcome; the factors we would reach for first if asked to make sense of why things unfolded the way they did.

Now, I have argued that data fusion and intelligence analysis are essential steps in the kill chain, and likely and often key determinants of why target X was hit rather than target Y. They are essential exactly because they are \textit{the primary drivers and key steps in explanations of outcomes and required steps in justification for why a particular outcome happened}. So if it is right that data fusion and intelligence analysis (and their outputs) are among the most important elements of explanations and justifications of morally salient outcomes (like striking targets or killing civilians) and if I am right that the relocation of initiative and interpretation happens when LLM-based agents are used for data fusion and intelligence analysis, then it follows that they cannot be used in a way that meets the condition of having context-appropriate human judgment or agency in the kill chain. In other words, the displacement of human judgment from ethically, normatively, and operationally most salient decision-making (data fusion and intelligence analysis) renders human presence in that part of the kill chain ineffectual in almost any sense and for almost any purpose.

\section*{V. Potential objections}

I take the above main claim to be rather intuitive given the normative and epistemic displacement of human judgment in LLM-based agents used for data fusion in the kill chain. After all, to normatively and epistemically displace a human \textit{just is} to render their judgment insignificant relative to the outcome (whether that outcome is a morally salient action, or a belief/judgment). To further support this claim, I turn now to potential objections.

First, a critic of the above conclusion might argue that the requirement for human judgment in the kill chain or targeting doesn't mean that a human must be present and fully making the decision in each step of the lifecycle of the weapon or kill chain.\footnote{Lifecycle is usually conceived of as development and design, TEVV, procurement, (maybe TEVV again), fielding and use. There are others who add post-deployment and/or training and post-deployment reflection (feedback loop) steps and pre-development political decisions, but the core of the lifecycle are the above steps.} After all, a critic might argue, we've got human judgment in the development of the agent and in TEVV (testing, evaluation, validation, and verification). Most of the policy processes mentioned above, the critic might continue, have already extended the meaning of ``human control'' to include pre-deployment, so why not think that human judgment in pre-deployment (specifically development and testing) is sufficient human control and sufficient human judgment.

In response, I think it is right to say that requirement of human judgment is not a requirement that every step of the way humans act as decision makers, but I also think that we would all agree that at some point the distance between human judgment/intent and the outcome is \textit{so mediated} and so far removed that such human judgment \textit{fails to meet} most of the \textit{purposes} it is meant to serve: safety, responsibility attribution, dignity, institutional stability. The features of goal-orientedness and sub-goal translation of human intent, together with the feature of dynamic memory, namely the ability to change the plan in response to changing contexts all lead to good reasons to think that neither developer's intent not commander's intent could sufficiently explain any outcome. That seems like a good reason to think that in LLM-based agent (serving data fusion and battlefield management) outcome is too far removed from human judgment. Obviously, what it means to have human judgment ``too far away'' from the outcome is one of the most important questions in this debate. However, if human intent (developers, commanders, or developers and commanders jointly) cannot in any meaningful way provide an explanation of a particular outcome, then we have good reason to think that the human judgment is not doing what those that insist on such judgment wanted in the first place.

Second, another potential objection is that if the humans are providing oversight over these systems (governance and other types of oversight), then that is a form of human judgment, and it might be sufficient to meet policy aims. This oversight could take the form of human-on-the-loop or just general real-time monitoring of agents' decisions and processes.

There are good reasons to think that true human oversight over these systems would satisfy at least some subset of those that insist on human judgment in the kill chain, so the above is a serious objection worth considering. But it is hard to imagine any real-time or even post-hoc oversight mechanism for agents that is not agentic itself. Between the stochastic nature of LLMs, dynamic memory and dynamic orchestration, alongside the enormous amount of processing and decision-making that goes into sifting through data at each step, real-time human oversight is not possible. Any human oversight would have to be mediated by another LLM agent, or another AI-tool that would also have to engage in interpretation itself. Whether this is a problem or not and whether humans providing oversight over agents providing oversight over LLM-based data fusion agents is sufficient, is an open question, and I do discuss it shortly in the final section.

Third potential objection is that the examples discussed in this paper (at least data fusion and intelligence analysis) are sufficiently upstream from outcomes and that ultimately, especially in cases of data fusion, a human gets presented with relevant evidence and information and makes a final decision. [4] Certainly there could be tools and times when that is not the case, in sort of traditional cases of autonomous weapons systems, but that is not our primary worry here. The primary worry here is that the use of LLM-based agents upstream (from outcomes) in data fusion and intelligence analysis make human judgment in that part of the kill chain ineffectual and vacuous and that that violates any and all reasons behind the insistence on context-appropriate human judgement in the kill chain. So, the critic here might argue that in most cases when it comes to final target engagement ultimately a human decides on the target or deployment of a weapon and that that is sufficient human judgment over the outcome that we are most interested in, namely the final strike/target selection and engagement. Isn't human making the final decision on targets sufficient human involvement in morally salient outcomes, the critic might conclude?

In response, there is good reason to think that this is not sufficient. Whatever the right ``amount'' or ``type'' of human judgement in the kill chain, the standard must be such that the process and content of that human judgment at least forms a part of the explanation of why a certain outcome occurred [12]. Ultimately it is not simply about human control in the physical sense, nor is it about humans simply being a but-for causes of the outcome, it is about human judgment in some meaningful way forming a part of an explanation of the outcome. Reasoning over a subset of data chosen by an LLM-based agent simply does not meet the conditions of human judgment in any substantive sense. Consider, what I take to be a similar case, where a police officer provides only some subset of information to a prosecutor, namely that someone resembling the accused was seen in the area, that the accused has the same type of a gun as the one that killed the victim, but he fails to include the fact that the accused has video-supported alibi for the time of the murder. Now imagine further that the prosecutor goes ahead based on the above evidence to charge the accused with the homicide. Months later we, the observers, find out that prosecutor was wrong, and that the defendant was innocent, and we want to know why she wrongly charged the innocent defendant. The key part of the explanation as to why this outcome happened is to be found in what data the police officer chose to include and exclude, not in the fact that prosecutor actually wrote and submitted the charging papers. Now, obviously there are times that with all the right information prosecutors charge people wrongly. But that is not what's at stake here, what's at stake here is what are the primary determinants of whether and why prosecutors (when acting correctly and in good faith) charge a potential defendant. The primary determinant in most cases is (or should be) the evidence that is given to the prosecutor. This seems to be very much the case on the battlefield as well. Why a target is struck, when it is struck, and how it is struck is determined by two main factors---the high-level intentions behind the operation or mission and the data that identifies and assesses the potential targets and the potential assets that could be used to engage that target. The processing of both of those main drivers of final battlefield outcomes (intentions and data) would in cases where LLM-based agents do data fusion and command and control functions be done by those agents rather than humans.\footnote{I want to stress another important response to both the second and third objections and that is that in both proposals---where human is providing some kind of real time oversight and where human is ultimately making the decision, the human thinking (judgment/deliberating) that would be occurring would be several layers removed from key facts---simply put, it would be such judgment is simply not about the same thing, it is not same in nature, and it would be performed by one kind of people (trained on assuring that agents are performing well, not making judgments or deliberation about the content of the judgments and decisions that agents are making, but about evidence that they are performing as expected). All of that would make those humans just not `close enough' to the outcome, ultimately performing (if any kind of judgment) judgment about different inputs. There are at least some who are perfectly comfortable with this and recommend instead that we actually agentify the agents. [11]}

\section*{VI. So what now?}

Given the above, what options are we left with? I think there are two obvious ones. One is to ban LLM-based agents (for the above capabilities), the other is to try to shift our thinking about human judgment in the kill chain. There are good reasons to think that banning the use of LLM-based agents in data fusion and for intelligence analysis is unrealistic. As I have argued in sections I and II, LLM-based agents are here, and they are here to stay, they are already being used in some forms on the battlefields and for these and similar purposes across a range of industries. This is not to say that it is not worth answering the questions whether in the ideal world they ought to be banned. In fact, if it turns out that all things considered they ought to be banned, then that might affect how we propose to regulate them (in a world where political expediency makes such a ban impossible). But centrally there are key steps we should undertake today, whether ultimately ideally these agents should be banned for these capabilities or not. These include:

\begin{enumerate}[label=\roman*.]
\item Resist overly relying on human judgment and control \textit{as a central part} of de-risking strategy for AI-enabled kill chains. This is especially important if there are parts of the kill chain where effectual human judgment or control is not possible. In those cases, reliance on human ``judgment'' and human ``involvement'' are window dressing that won't serve the purpose for which such judgment was intended and will in fact detract from achieving the main aims of our policies, because insisting on human judgment (in such cases) risks distracting our focus from other de-risking strategies.

\item Develop new methods for testing, evaluating, validating, and verifying these systems (TEVV), including red teaming, and real world TEVV, as well as TEVV with human-machine teams.

\item Develop and require significant oversight mechanisms for LLM-based agents including counterfactual real-time processing, guardrails, elements of orchestration layers that do not get interpreted by LLMs, and similar.

\item Finally, in as much as we want to or can retain some human judgment as a part of the solution---either in other parts of the kill chain or several steps removed from data fusion (both vertically and horizontally), we should focus on setting limits for \textit{how far removed human judgment} can be from the outcomes, which should depend on:

\begin{enumerate}[label=\alph*.]
\item why we wanted human judgment in the first place (is it safety, or responsibility attribution, or dignity, and or institutional stability); and
\item ordinary language analysis and legal analysis of how we attribute causation for/explanations of outcomes; and
\item recognize that in some cases human judgment will not be about things that drive the outcomes directly. In other words, if we have governance AI agents providing oversight over LLM agents for data fusion, the human oversight over the governance agents is about other things and is not human oversight over determinants of battlefield outcomes. This will require very different training and responsibility attribution frameworks.
\end{enumerate}
\end{enumerate}

All in all, if the above is right, continued insistence on human judgment is detracting from genuine and direct solutions for a lot of the main worries, namely those about safety, loss of control, responsibility attribution, escalation, and similar. AI tools for warfighting, including LLM-based agents used for data fusion and similar tasks, present a range of serious risks: to the safety of civilians and soldiers, to global stability in the face of escalation, and to meaningful responsibility assignment. Some of those risks can be meaningfully alleviated through human engagement, control, and judgment (for example, safety risks arising from object recognition AI systems used for target identification or selection on one-way effectors). But in cases where LLM-based agents are used for data fusion and when they displace human judgment from the relevant causal story behind an outcome, insisting on nominal human involvement, in the form of oversight or final sign-off, is little more than window-dressing, and will not contribute to safety or any of the other key aims in any substantive way.

Agents and LLMs are coding, they are building simulations for testing, they are used for compliance checks in procurement, they are used for training, and for some battlefield function (data fusion, targeting, etc.) [5]. When this becomes the norm, as opposed to an exception, we need to have a ready answer to the question what human engagement in this entire lifecycle looks like and why it matters. For now, there are good reasons to worry that human judgment is rendered ineffectual and meaningless when LLM-based agents are used for data fusion and battlefield management.

\section*{Acknowledgements}

This project is sponsored by the Research Council of Norway 3-year research grant ``Ethical Risk Management of AI-enabled weapons systems,'' grant number 352870. The author is grateful to audiences at the 2026 REAIM Summit, as well as the Peace Research Institute Oslo research team (on the above grant) and the BABL AI team who have all provided meaningful feedback.

\section*{References}

\begin{enumerate}[label={[{\arabic*}]}, leftmargin=*, itemsep=2pt]

\item Anduril. 2026. \textit{Lattice Command and Control}. \url{https://www.anduril.com/lattice/command-and-control}. Accessed April 2026.

\item {The AutoPractices Project}. 2026. \textit{Strengthening Human Agency in the Military Domain: Best Practices Toolkit for Policymakers, Developers, and Users of AI Systems}. Center for War Studies, University of Southern Denmark, Odense. \url{https://www-static.autonorms.eu/wp-content/uploads/2026/01/AutoPractices-toolkit-WEB.pdf}. Accessed April 2026.

\item Julian E. Barnes, Eric Schmitt, Tyler Pager, Malachy Browne, and Helene Cooper. 2026. \textit{U.S. at Fault in Strike on School in Iran, Preliminary Investigation Finds}. The New York Times. \url{https://www.nytimes.com/2026/03/11/us/politics/iran-school-missile-strike.html}. Accessed April 2026.

\item Ingvild Bode. 2025. \textit{Human-Machine Interaction and Human Agency in the Military Domain}. Policy Brief No.\ 193. Centre for International Governance Innovation, Waterloo, ON. \url{https://www.cigionline.org/static/documents/PB_no.193.pdf}. Accessed April 2026.

\item Stephen Casper, Luke Bailey, Rosco Hunter, Carson Ezell, Emma Cabal\'{e}, Michael Gerovitch, Stewart Slocum, Kevin Wei, Nikola Jurkovic, Ariba Khan, Phillip Christoffersen, Pinar Ozisik, Rakshit Trivedi, Dylan Hadfield-Menell, and Noam Kolt. 2025. \textit{AI Agent Index: Documenting the Technical and Safety Features of Deployed Agentic AI Systems}. MIT. \url{https://aiagentindex.mit.edu/}. Accessed April 2026.

\item M.L. Cummings. 2025. \textit{Prohibiting Generative AI in any Form of Weapon Control}. In \textit{39th Conference on Neural Information Processing Systems (NeurIPS 2025) Position Paper Track}. \url{https://openreview.net/pdf?id=uEY7kQsiZz}.

\item {CCW Group of Governmental Experts on Lethal Autonomous Weapons Systems}. 2025. \textit{GGE on LAWS Rolling Text, Status Date: 18 December 2025}. Convention on Certain Conventional Weapons, United Nations Office for Disarmament Affairs. \url{https://docs-library.unoda.org/Convention_on_Certain_Conventional_Weapons_-Group_of_Governmental_Experts_on_Lethal_Autonomous_Weapons_Systems_(2026)/CCW_GGE_LAWS_Rolling_Text_-_status_18_December_2025.pdf}. Accessed April 2026.

\item Jovana Davidovic. 2025. Rethinking Human Roles in AI Warfare. \textit{Nature Machine Intelligence} 7 (2025), 1593--1595. \url{https://doi.org/10.1038/s42256-025-01123-6}.

\item Jovana Davidovic. 2023. On the Purpose of Meaningful Human Control of AI. \textit{Frontiers in Big Data} 5 (2023), 1017677. \url{https://doi.org/10.3389/fdata.2022.1017677}.

\item {Defense Innovation Unit}. 2026. \textit{DIU and DAWG Launch Autonomous Vehicle Orchestrator Prize Challenge}. Defense Innovation Unit. \url{https://www.diu.mil/latest/diu-and-dawg-launch-autonomous-vehicle-orchestrator-prize-challenge}. Accessed April 2026.

\item Virginia Dignum and Frank Dignum. 2025. \textit{Agentifying Agentic AI}. arXiv:2511.17332. \url{https://doi.org/10.48550/arXiv.2511.17332}.

\item Madeleine Clare Elish. 2019. Moral Crumple Zones: Cautionary Tales in Human-Robot Interaction. \textit{Engaging Science, Technology, and Society} 5 (2019). \url{https://doi.org/10.17351/ests2019.260}.

\item {Global Commission on Responsible Artificial Intelligence in the Military Domain}. 2025. \textit{Responsible by Design: Strategic Guidance Report on the Risks, Opportunities, and Governance of Artificial Intelligence in the Military Domain}. The Hague Centre for Strategic Studies, The Hague.

\item Lewis Hammond, Alan Chan, Jesse Clifton, Jason Hoelscher-Obermaier, Akbir Khan, Euan McLean, Chandler Smith, et al. 2025. \textit{Multi-Agent Risks from Advanced AI}. Cooperative AI Foundation Technical Report \#1. arXiv:2502.14143. \url{https://doi.org/10.48550/arXiv.2502.14143}.

\item Michael C. Horowitz and Lauren Kahn. 2026. \textit{First Ukraine, Now Iran: A New Era of Drone Warfare Takes Hold}. Council on Foreign Relations. \url{https://www.cfr.org/articles/the-new-era-of-drone-warfare-takes-root-in-iran}. Accessed April 2026.

\item Atoosa Kasirzadeh and Iason Gabriel. 2025. \textit{Characterizing AI Agents for Alignment and Governance}. arXiv:2504.21848. \url{https://doi.org/10.48550/arXiv.2504.21848}.

\item Anne Kleppe, Steven Mills, Noah Broestl, Grigor Acenov, Kirill Katsov, and Ning Yang. 2025. \textit{What Happens When AI Stops Asking Permission?} Boston Consulting Group. \url{https://www.bcg.com/publications/2025/what-happens-ai-stops-asking-permission}. Accessed April 2026.

\item Courtney Kube. 2026. \textit{Democrats Ask Pentagon About Iran School Strike and Role of AI}. NBC News. \url{https://www.nbcnews.com/politics/national-security/democrats-ask-pentagon-iran-school-strike-role-ai-rcna263083}. Accessed April 2026.

\item Elena Marchetti. 2026. \textit{AI Likely Caused Iran School Bombing That Killed 175}. Awesome Agents. \url{https://awesomeagents.ai/news/ai-error-iran-school-bombing-minab/}. Accessed April 2026.

\item {NATO}. 2024. \textit{Summary of NATO's Revised Artificial Intelligence (AI) Strategy}. NATO. \url{https://www.nato.int/en/about-us/official-texts-and-resources/official-texts/2024/07/10/summary-of-natos-revised-artificial-intelligence-ai-strategy}. Accessed April 2026.

\item David Szondy. 2026. \textit{Watch: Fury Platform Brings Agentic AI to Battlefield Drones}. New Atlas. \url{https://newatlas.com/military/video-fury-ai-platform-battlefield-drones/}. Accessed April 2026.

\item {U.S. Department of Defense}. 2023. \textit{Directive 3000.09: Autonomy in Weapon Systems}. DoD Directive 3000.09.

\item {U.S. Department of War}. 2026. \textit{Artificial Intelligence Strategy for the Department of War: Accelerating America's Military AI Dominance}. U.S. Department of Defense. \url{https://media.defense.gov/2026/Jan/12/2003855671/-1/-1/0/ARTIFICIAL-INTELLIGENCE-STRATEGY-FOR-THE-DEPARTMENT-OF-WAR.PDF}. Accessed April 2026.

\item Julia Wiesinger, Patrick Marlow, and Vladimir Vuskovic. 2024. \textit{Agents}. Google/Kaggle. \url{https://www.kaggle.com/whitepaper-agents}. Accessed April 2026.

\item Yasaman Yousefi, Marco Billi, and Antonino Rotolo. 2025. \textit{Agentic AI: An EU AI Act Paradigm Shift?} SSRN. \url{https://papers.ssrn.com/sol3/papers.cfm?abstract_id=5731424}. Accessed April 2026.

\end{enumerate}

\end{document}